\documentclass[aps,showpacs,twocolumn,prl,superscriptaddress]{revtex4-1}
\usepackage{amsmath}
\usepackage{amssymb}
\usepackage{mathtext}
\usepackage{graphicx}
\usepackage{color}

\begin{document}

\title{
Dynamics of noisy oscillator populations beyond the Ott-Antonsen ansatz}
\author{Irina V.\ Tyulkina}
\affiliation{Department of Theoretical Physics, Perm State University, Bukirev Street 15,
 Perm 614990, Russia}
\author{Denis S.\ Goldobin}
\affiliation{Department of Theoretical Physics, Perm State University, Bukirev Street 15,
 Perm 614990, Russia}
\affiliation{Institute of Continuous Media Mechanics, UB RAS, Academician Korolev Street 1,
 614013 Perm, Russia}
\author{Lyudmila S.\ Klimenko}
\affiliation{Department of Theoretical Physics, Perm State University, Bukirev Street 15,
 Perm 614990, Russia}
\affiliation{Institute of Continuous Media Mechanics, UB RAS, Academician Korolev Street 1,
 614013 Perm, Russia}
\author{Arkady Pikovsky}
\affiliation{Institute for Physics and Astronomy,
University of Potsdam, Karl-Liebknecht-Strasse 24/25, 14476 Potsdam-Golm, Germany}
\affiliation{Research Institute for Supercomputing, Nizhny Novgorod State University,
Gagarin Avenue 23, 606950 Nizhny Novgorod, Russia}
\date{\today}

\begin{abstract}
We develop an approach for the description of the dynamics of large populations of phase oscillators based on ``circular cumulants'' instead of the Kuramoto-Daido order parameters. In the thermodynamic limit, these variables yield a simple representation of the Ott-Antonsen invariant solution [E.\ Ott and T.\ M.\ Antonsen, CHAOS {\bf 18}, 037113 (2008)] and appear appropriate for constructing the perturbation theory on top of the Ott-Antonsen ansatz.
We employ this approach to study the impact of small intrinsic noise on the dynamics. As a result, a closed system of equations for the two leading cumulants, describing the dynamics of noisy ensembles, is derived. We exemplify the general theory by presenting the effect of noise on the Kuramoto system and on a chimera state in two symmetrically coupled populations.
\end{abstract}

\maketitle

The study of ensembles of self-sustained (autonomous) oscillators is relevant for many setups in physics (Josephson junction and spin-torque oscillator arrays~\cite{Benz-Burroughs-91,*Pikovsky-13,*Turtle_etal-17}), engineering (stability of pedestrian bridges, coupled electronic oscillators, power grid networks~\cite{Dallard-01a,*Temirbayev_et_al-12,*Filatrella_etall-08}), chemistry (ensembles of electrochemical oscillators~\cite{Kiss-Zhai-Hudson-02a}), and life sciences (colonies of yeast cells, synthetic gene oscillators~\cite{Richard-Bakker-Teusink-Van-Dam-Westerhoff-96,*Prindle_etal-12}). Of major interest are collective phenomena, like synchronization, in these ensembles. Many essential properties can be described already within simple phase models valid at small coupling. In this limit, only the dynamics of the phases of oscillators is nontrivial, while the amplitudes are enslaved. The paradigmatic model here is the Kuramoto model of sine-coupled phase oscillators~\cite{Kuramoto-75}, demonstrating a nonequilibrium transition from asynchrony to synchrony~\cite{Kuramoto-84}.

Recently, remarkable progress has been achieved in the description of the dynamics of order parameters for populations of phase oscillators. Watanabe and Strogatz (WS)~\cite{Watanabe-Strogatz-93,*Watanabe-Strogatz-94} demonstrated partial integrability of a class of phase ensembles, allowing reduction of the full dynamics of a population of $N$ identical elements to a three-dimensional set of equations for certain global variables, plus $N-3$ constants of motion.
In the thermodynamic limit, where the number of the constants of motion tends to infinity, this integrability means invariance in time of the density of the constants. The global variables have an especially transparent form if the density of the constants is uniform---in this case one obtains a closed dynamical system for the natural order parameters of the system. These equations have been obtained by Ott and Antonsen (OA)~\cite{Ott-Antonsen-08} with another method, see Refs.~\cite{Pikovsky-Rosenblum-08,Marvel-Mirollo-Strogatz-09} for interrelation of the two approaches. In the full state space, the OA equations are valid on a particular OA manifold, which is only neutrally stable for identical oscillators (due to WS integrability), but becomes weakly stable if one performs coarse graining for nonidentical oscillators~\cite{Ott-Antonsen-09,*Mirollo-12,*Pietras-Daffertshofer-16}. The simplicity of OA equations made them a popular tool in studies of many setups like coupled ensembles~\cite{Martens_etal-09,*Komarov-Pikovsky-11,*So-Barreto-11,*Komarov-Pikovsky-13}, chimera states~\cite{Abrams-Mirollo-Strogatz-Wiley-08,Laing-09,*Bordyugov-Pikovsky-Rosenblum-10} consisting of synchronized and partially synchronous parts, common-noise driven~\cite{Nagai-Kori-10,*Braun-etal-12}, excitable, and non-homogeneous phase oscillators~\cite{Laing-12,Laing-14,*Luke-Barreto-So-14,*Montbrio-Pazo-Roxin-15}.

The main goal of this letter is to extend the OA theory to the case of noisy oscillators. For small noise, in the leading order in noise intensity, we will derive and analyze a closed dynamical system for the two order parameters, describing populations of noisy phase oscillators. Our main tool is the reformulation of the full dynamics in terms of the circular cumulants, which are related to the complex order parameters in the same way that cumulants of distributions of real random variables are related to their moments. We stress here, that because the complex order parameters are moments of a complex observable defined on the unit circle, circular cumulants have nothing in common with Gaussian approximations sometimes used in theories of collective dynamics~\cite{Zaks-etal-03}.

We start with an ensemble of phase oscillators $\varphi_k(t)$ having the same natural frequency $\Omega$, subject to a common complex-valued external force $h(t)$, and to intrinsic (not common) noise:
 \begin{equation}
\dot\varphi_k=\Omega+\mathrm{Im}(2h(t)e^{-i\varphi_k})+{\sigma}{\xi_k(t)}\,,
\label{eq101}
\end{equation}
where $\xi_k$ are independent white Gaussian noises: $\langle\xi_k(t)\rangle=0$, $\langle\xi_k(t) \xi_m(t')\rangle=2\delta_{km} \delta(t-t')$. In most  applications the force itself depends  on the phases via mean-field coupling, but to develop the theory it is convenient to write it in a general form. In the thermodynamic limit of an infinite ensemble, its state can be described by the distribution density $w(\varphi,t)$ which obeys the Fokker-Planck equation
\begin{equation}
\frac{\partial w}{\partial t}
 +\frac{\partial}{\partial\varphi}((\Omega-ihe^{-i\varphi}+ih^\ast e^{i\varphi})w)
 ={\sigma^2}\frac{\partial^2 w}{\partial\varphi^2}\,.
\label{eq102}
\end{equation}
It is convenient to introduce Fourier modes,
$w(\varphi,t)=(2\pi)^{-1}[a_0+\sum_{j=1}^{\infty}(a_je^{-ij\varphi}+c.c.)]$, with $a_0=1$ due to the normalization condition, and to write an infinite system of equations for their evolution (cf.~\cite{Laing-12})
\begin{equation}
\dot{a}_{j}=ji\Omega a_j+jh{a}_{j-1}-jh^\ast{a}_{j+1}
 -{\sigma^2}j^2{a}_j\,,\quad j\geq 1\;.
\label{eq113}
\end{equation}
Complex quantities $a_j=\langle e^{i j\varphi}\rangle$ are nothing else but the Kuramoto-Daido order
parameters~\cite{Daido-96} for the ensemble.

In the case of a population of non-identical oscillators with different natural frequencies $\Omega$ (we assume that the forcing $h$ is still a common one), one can consider the Fourier modes $a_j(\Omega,t)$ as functions of $\Omega$. It is natural to introduce order parameters averaged over frequencies $\Omega$ with the distribution $g(\Omega)$:
$Z_j=\int g(\Omega)\,a_j(\Omega,t)\,d\Omega$. This averaging takes an especially simple form for the Lorentzian distribution of frequencies $g(\Omega)=\frac{\gamma}{\pi((\Omega-\Omega_0)^2+\gamma^2)}$, where $\gamma$  is the characteristic half-width of the natural frequency band around the central frequency $\Omega_0$. With the assumption that all $a_j(\Omega,t)$ are analytic in the upper half-plane of the complex $\Omega$-plane (if they are analytic at certain time instant $t_\ast$, they will remain analytic for $t>t_\ast$; see \cite{Ott-Antonsen-08} for details), the integral can be evaluated via residues: $Z_j(t)=a_j(\Omega_0+i\gamma,t)$. As a result, we obtain an infinite system of equations for the order parameters $Z_j$:
\begin{equation}
\dot{Z}_{j}=j(i\Omega_0 -\gamma) Z_j+jh{Z}_{j-1}-jh^\ast{Z}_{j+1}
 -{\sigma^2}j^2{Z}_j\,.
\label{eq124}
\end{equation}
We stress here, that the analyticity property is important only for the reduction of the sets of the Fourier modes to single values calculated at a pole in the complex plane. For an ensemble of identical oscillators (which formally corresponds to $\gamma=0$) and for frequency distributions which cannot be integrated via residues, analyticity is irrelevant, and our method below does not rely on this property.

In the noise-free case $\sigma=0$, the system \eqref{eq124} possesses an invariant two-dimensional manifold, called Ott-Antonsen manifold~\cite{Ott-Antonsen-08}: $Z_n=(Z_1)^n$. The resulting equation for the main order parameter $Z_1$ has a simple form
\begin{equation}
\dot{Z}_1=(i\Omega_0 -\gamma) Z_1+h-h^\ast{Z}_{1}^2\;.
\label{eqOA}
\end{equation}
On the OA manifold, all the order parameters are determined by $Z_1$ and hence $h$ is a function of $Z_1$; therefore, Eq.~\eqref{eqOA} is a closed equation, the dynamics of which is easy to analyze. This made the OA ansatz so popular in different setups.

Our goal in this letter is to develop a low-dimensional description of the dynamics of the order parameters in the presence of weak noise $\sigma\neq 0$. The main idea is to reformulate the general equations \eqref{eq124} in a cumulant form, more suitable for the perturbation approach.

The order parameters $Z_n=\langle e^{in\varphi}\rangle$ can be treated as moments of the observable $e^{i\varphi}$, and they can be obtained from the moment-generating function
\begin{equation}
F(k,t)=
\langle\exp(ke^{i\varphi})\rangle
\equiv\sum_{m=0}^{\infty}Z_m(t)\frac{k^m}{m!}
\label{eq201}
\end{equation}
as $Z_m(t)=\frac{\partial^m}{\partial k^m}F(k,t)|_{k=0}$. The partial differential equation for $F$ follows directly from Eq.~\eqref{eq124}:
 \begin{equation}
 \begin{aligned}
\frac{\partial F}{\partial t}=&(i\Omega_0-\gamma) k\frac{\partial}{\partial k}F
 +hkF -h^\ast k\frac{\partial^2}{\partial k^2}F\\&
 -\sigma^2k\frac{\partial}{\partial k}\left(k\frac{\partial}{\partial k}F\right).
 \end{aligned}
\label{der01}
\end{equation}
As in many other situations, it appears beneficial to introduce \textit{circular cumulants} $\varkappa_m$
via the power series of the cumulant-generating function defined as~\cite{footnote1}
\begin{equation}
\varPsi(k,t)=
k\frac{\partial}{\partial k} \ln F(k,t)=
\frac{k}{F}\frac{\partial F}{\partial k}
\equiv
\sum_{m=1}^{\infty}\varkappa_{m}(t)\,k^m\,.
\label{eq:cce}
\end{equation}
For example, the first three circular cumulants are:
\[
\varkappa_1=Z_1,\quad\varkappa_2=Z_2-Z_1^2,\quad \varkappa_3=\frac{1}{2}(Z_3-3Z_2Z_1+2Z_1^3)\;.
\]
The partial differential equation for $\varPsi(k,t)$ can be derived by applying the operator $\partial_t$ to \eqref{eq:cce}, and employing \eqref{der01}:
\begin{align}
\frac{\partial\varPsi}{\partial t}&
=(i\Omega_0-\gamma)k\frac{\partial\varPsi}{\partial k}
 +hk
 -h^\ast k\frac{\partial}{\partial k}
 \left(k\frac{\partial}{\partial k}\left(\frac{\varPsi}{k}\right)
 +\frac{\varPsi^2}{k}\right)
\nonumber\\
&
\qquad
 -\sigma^2k\frac{\partial}{\partial k}\left(k\frac{\partial\varPsi}{\partial k}+\varPsi^2\right).
\label{der03}
\end{align}
The infinite system of equations for the circular cumulants
can be obtained directly from (\ref{eq:cce},\ref{der03}):
\begin{equation}
\begin{aligned}
\dot\varkappa_j&=j(i\Omega_0 -\gamma)\varkappa_{j}+h\delta_{j1}
\\
& -h^\ast(j^2\varkappa_{j+1}+j\sum_{m=1}^{j}\varkappa_{j-m+1}\varkappa_{m})
\\&
\quad
-{\sigma^2}(j^{2}{\varkappa}_{j}+j\sum_{m=1}^{j-1}\varkappa_{j-m}\varkappa_{m})\,.
\end{aligned}
\label{der05}
\end{equation}

The advantage of circular cumulants is in a simple representation of the Ott-Antonsen manifold:
It corresponds to the case where all high cumulants vanish: $\varkappa_m=0,\;m>1$; and the only nontrivial cumulant is $\varkappa_1=Z_1$. In this case the generating functions are $\varPsi(k,t)=k Z_1(t)$ and $F(k,t)=\exp[k Z_1]$. One can easily check that these generating functions are invariant solutions of Eqs.~\eqref{der01} and \eqref{der03} for vanishing noise $\sigma=0$, provided $Z_1$ evolves according to Eq.~\eqref{eqOA}.

For non-vanishing noise, generally all the cumulants are non-zero. However, for small noise one can expect the cumulants with orders larger than one to be small. To reveal the hierarchy of this smallness, it is instructive to write explicitly the equations for the cumulants $\varkappa_2$ and $\varkappa_3$:
\begin{align}
\dot\varkappa_2&=2i(\Omega_0-\gamma)\varkappa_2-4 h^\ast \varkappa_3-4 h^\ast \varkappa_1\varkappa_2\nonumber\\
&\quad-4\sigma^2 \varkappa_2
-2\sigma^2\varkappa_1^2\;,
\label{eq:c2}\\
\dot\varkappa_3&=3i(\Omega_0-\gamma)\varkappa_3-9h^\ast \varkappa_4-
h^* [6\varkappa_1\varkappa_3+3\varkappa_2^2]\nonumber\\
&\quad-9\sigma^2 \varkappa_3 -6\sigma^2 \varkappa_1\varkappa_2\;.\label{eq:c3}
\end{align}
On the r.h.s.\ of Eq.~\eqref{eq:c2} there are ``homogeneous'' terms $\propto\varkappa_2$ and ``driving'' terms $\sim\varkappa_3$ and $\sim\sigma^2\varkappa_1^2$. If we assume that the higher-order cumulants are smaller than the lower-order ones, then the term $\sim\sigma^2\varkappa_1^2$ determines the level of the cumulant $\varkappa_2$, which appears to be $\varkappa_2\sim\sigma^2\varkappa_1^2$. A similar inspection of Eq.~\eqref{eq:c3} yields leading ``driving'' terms $\sim \varkappa_2^2$ and $\sim\sigma^2\varkappa_1\varkappa_2$, both have order $\sim\sigma^4$. Thus we conclude that the smallness of the third cumulant is $\sim\sigma^4$. Inspection of the full system~\eqref{der05} shows that an assumption $|\varkappa_m|\sim\sigma^{2(m-1)}$ is consistent with the dynamics in all orders. A more detailed analysis of the hierarchy of cumulants will be reported elsewhere; here below we will exploit the simplest approximation, where  we assume all the cumulants above the second one to vanish. As it follows from the above discussion, accuracy of this approximation is $\mathcal{O}(\sigma^4)$. As a result, we obtain a closed system of equations for the first and the second cumulants (for simplicity of further notations, we omit the index of $Z_1$ and denote $\kappa=\varkappa_2$):
\begin{equation}
\begin{aligned}
\dot{Z}&=(i\Omega_0-\gamma)Z+h-h^* Z^2-\sigma^2 Z-
h^*\kappa\;,\\
\dot{\kappa}&=2(i\Omega_0-\gamma)\kappa-4h^* Z\kappa-\sigma^2
(4\kappa+2Z^2)\;.
\end{aligned}
\label{eq2c}
\end{equation}
This system of two equations for two complex order parameters $Z$ and $\kappa=Z_2-Z_1^2$ generalizes the Ott-Antonsen equation~\eqref{eqOA} to the case of small noise. Below we will explore it in different setups.

It is instructive to examine the perturbation of the OA probability density corresponding to the one nonvanishing second circular cumulant $\kappa$. With two nonvanishing cumulants, the moment-generating function is $F(k)=\exp\big[kZ+\kappa\frac{k^2}{2}\big]$. Assuming smallness of $\kappa$, we approximate it as $F(k)\approx (1+\kappa\frac{k^2}{2})\exp[kZ]$, and obtain for the moments $Z_m=Z^m+\frac{m(m-1)}{2}\kappa Z^{m-2}$. Summation of the Fourier series with these Fourier coefficients yields $w(\varphi)=w_{OA}(\varphi)+w_C(\varphi)$, where
\begin{equation}
w_{OA}(\varphi)=\frac{1-|Z|^2}{2\pi|e^{i\varphi}-Z|^2}\,,\quad
w_C(\varphi)=\text{Re}\!\left[\frac{\pi^{-1}\kappa e^{i\varphi}}{\left(e^{i\varphi}-Z\right)^3}\right].
\nonumber
\end{equation}
The perturbation $w_C$ is a function of the relative phase $\varphi-\text{arg}(Z)$ and depends on the parameter $\Theta=\text{arg}(\kappa)-2\text{arg}(Z)$.

\begin{figure}[!thb]
\centerline{
\includegraphics[width=0.97\columnwidth]%
 {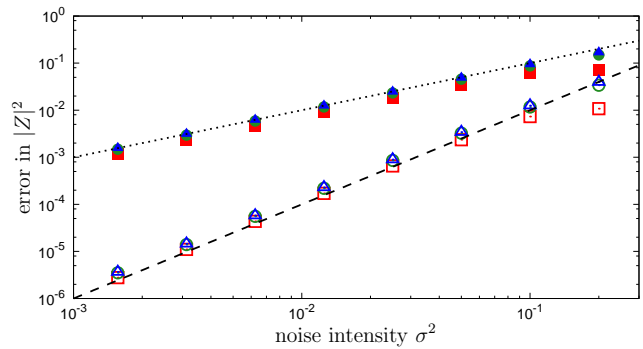}
}
\caption{(color online) Error of the approximate solution~\eqref{eq:Kss} as a
function of $\sigma^2$, for $\varepsilon=1$ and
different values of parameter $\gamma$: red open squares $\gamma=0.2$; green open
circles $\gamma=0.05$; blue open triangles $\gamma=0$. Dashed line shows theoretical
prediction  $\sim\sigma^4$. We show also
predictions of the simplified model where one sets $\kappa=0$,
with the corresponding filled markers. The error of this approximation is $\sim\sigma^2$ (dotted line).}
  \label{fig1}
\end{figure}

As a first application of the theory, we consider the effect of noise on the standard Kuramoto problem, where the ensemble is driven by the mean field itself, i.e.\ $h=\frac{\varepsilon}{2}Z$, where $\varepsilon$ is the coupling constant. By a transformation to a rotating with frequency $\Omega_0$ reference frame we can set $\Omega_0=0$. Now the system~\eqref{eq2c} takes the form
\begin{equation}
\begin{aligned}
\dot{Z}&=-\gamma Z+\frac{\varepsilon}{2}Z(1-|Z|^2) -\sigma^2Z-\frac{\varepsilon}{2}Z^\ast\kappa\;,\\
\dot{\kappa}&=-2\gamma\kappa-2\varepsilon|Z|^2\kappa -\sigma^2(4\kappa+2Z^2)\;.
\end{aligned}
\label{eq2cK}
\end{equation}
The dynamics of this model is simple: above the instability threshold of the asynchronous state $Z=\kappa=0$, which is $\varepsilon_c=2(\gamma+\sigma^2)$, the system evolves to a stable steady state
\begin{equation}
\begin{aligned}
|Z|^2&=\frac{1}{2}-\frac{3(\gamma+\sigma^2)}{2\varepsilon}\\
&\quad+\frac{\sqrt{(\varepsilon-\gamma)^2+2\sigma^2(\varepsilon-3\gamma)-7\sigma^4}}{2\varepsilon}\;,\\
\kappa&=\frac{-\sigma^2 Z^2}{\gamma+2\sigma^2+\varepsilon|Z|^2} \;.
\end{aligned}
\label{eq:Kss}
\end{equation}
Comparison of this solution with the numerical solution of full Eqs.~\eqref{eq124} in Fig.~\ref{fig1} shows that indeed the approximation \eqref{eq2cK} has accuracy $\sim\sigma^4$ in the whole range of $\gamma$, including the case of identical oscillators $\gamma=0$.

In this figure we also show predictions of the simplest approximation, where one sets all higher circular cumulants, except the first one, to zero. In this approximation, the system~\eqref{eq2c} reduces to one simple equation
$\dot Z=(i\Omega_0-\gamma)Z+h-h^* Z^2-\sigma^2 Z$.
The solution of the Kuramoto model then reads
$|Z|^2=(\varepsilon-2\gamma-2\sigma^2)/\varepsilon$. As it follows from comparison with solution~\eqref{eq:Kss}, the error of this approximation is of the order $\sim\sigma^2$; this is confirmed by calculations presented in Fig.~\ref{fig1}. Thus, this approximation, although it leads to very simple equations, does not catch the effect of noise in the leading order, contrary to the system~\eqref{eq2c}.

\begin{figure}[!thb]
\centering
\includegraphics[width=\columnwidth]%
 {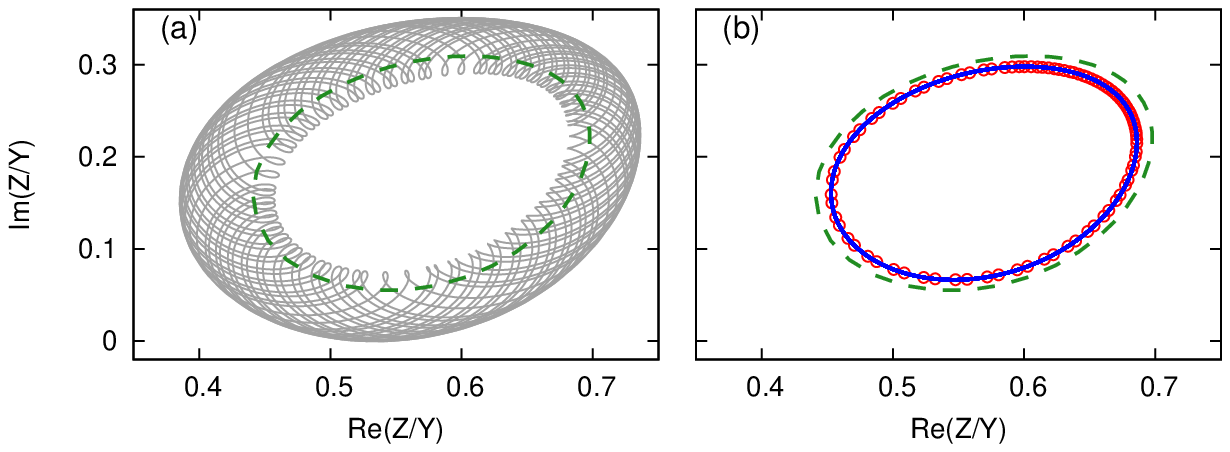}
\includegraphics[width=0.85\columnwidth]%
 {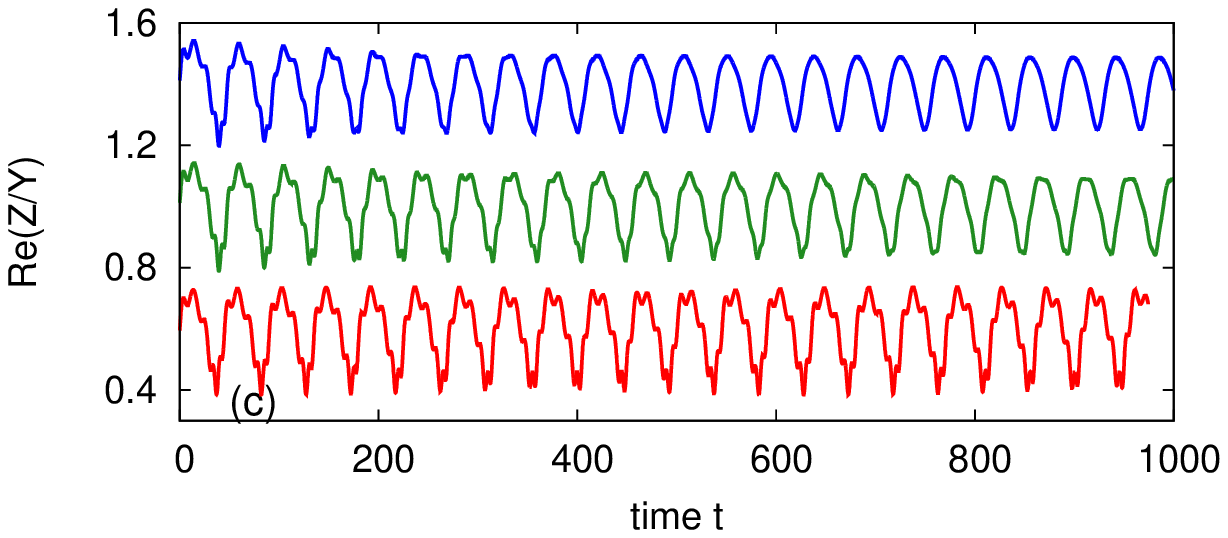}
 \caption{(color online) Panel (a): Noise-free chimera dynamics.
 Dashed green line: periodic solution  on the OA manifold; solid grey line: quasiperiodic solution out of the OA manifold.
 Panel (b): Dynamics in the presence of noise $\sigma^2=10^{-4}$. Solid blue line (solution of the full equations \eqref{eq113}) is overlapped by red circles (solution of system~\eqref{eq:abr2}). Dashed green line is the same as in panel (a). Panel (c): time evolution of the order parameters in the noise-free case (bottom red line, this solution corresponds to the grey solid line in panel (a)), for $\sigma^2=4\cdot 10^{-5}$ (middle green line), and for $\sigma^2=10^{-4}$ (top blue line). Middle and top lines are shifted for better visibility. All solutions start from the same initial conditions out of the OA manifold.
 }
  \label{fig2}
\end{figure}

With the next example we illustrate that small noise acts as a factor, stabilizing a vicinity of the OA manifold in systems of identical oscillators. The system suggested by Abrams et al.~\cite{Abrams-Mirollo-Strogatz-Wiley-08} consists of two symmetrically coupled populations (variables $\varphi$ and $\psi$) of phase oscillators. We write this system with additional independent white noise terms:
\begin{equation}
\begin{aligned}
\dot\varphi_k=&\Omega+\frac{1+A}{2N}\sum _{j=1}^N\sin(\varphi_j-\varphi_k-\alpha)\\
&+\frac{1-A}{2N}\sum _{j=1}^N\sin(\psi_j-\varphi_k-\alpha) +\sigma\xi_k(t)\;,\\
\dot\psi_k=&\Omega+\frac{1+A}{2N}\sum _{j=1}^N\sin(\psi_j-\psi_k-\alpha)\\
&+\frac{1-A}{2N}\sum _{j=1}^N\sin(\varphi_j-\psi_k-\alpha) +\sigma\eta_k(t)\;.
\end{aligned}
\label{eq:abr1}
\end{equation}
Here $N$ is the size of the populations, $\alpha$ is the phase shift in the coupling. Parameter $A$ determines different coupling strengths of intra- and inter-population interactions. Noise-free ($\sigma=0$) regimes have been analyzed in Refs.~\cite{Abrams-Mirollo-Strogatz-Wiley-08,Pikovsky-Rosenblum-08}; for experimental realization of this setup see \cite{Tinsley_etal-12,Martens_etal-13}. In Ref.~\cite{Abrams-Mirollo-Strogatz-Wiley-08} it was shown, using the OA ansatz, that in a range of parameters a regime where one population fully synchronizes (i.e., $\psi_1=\ldots=\psi_N=\Psi$), while the other one is partially synchronous (i.e., its order parameter $Z=\langle e^{i\varphi}\rangle$ takes values $0<|Z|<1$), is stable. This chimera state in the reference frame rotating with the phase of the second population $\Psi$, can be static (i.e., $Ze^{-i\Psi}=\text{const}$) or periodic (i.e., $Ze^{-i\Psi}$ is a periodic function of time). In Ref.~\cite{Pikovsky-Rosenblum-08} it was shown that the regimes studied in~\cite{Abrams-Mirollo-Strogatz-Wiley-08} are observed only if the initial conditions lie on the OA manifold for the first population. Because the OA manifold is not attractive, for more general initial conditions, one more nontrivial frequency is added: one observes a periodic regime instead of a steady state, and a quasiperiodic regime instead of a periodic one. This is illustrated in Fig.~\ref{fig2}(a). The dashed green line shows a periodic solution on the OA manifold, while the solid grey line shows a quasiperiodic regime for the initial conditions away from the OA manifold.
The bifurcation analysis for the attracting regimes in system~\eqref{eq:abr1} with intrinsic noise, performed in the thermodynamic limit $N\to\infty$ within the framework of Eqs.~(\ref{eq113}), can be also found in~\cite{Laing-12}.

We now apply to system~\eqref{eq:abr1} the small-noise theory developed above. In the presence of noise both populations are partially synchronous, thus we have to write a system of two equations of type~\eqref{eq2c} for two order parameters $Z$, $Y$ and for two corresponding second cumulants $\kappa$, $\nu$. Here also enter two fields acting on the populations, $H=0.25((1+A)Z+(1-A)Y)e^{-i\alpha}$ and $F=0.25((1+A)Y+(1-A)Z)e^{-i\alpha}$ (we set $\Omega=0$, because this parameter can be excluded in the rotating reference frame):
\begin{equation}
\begin{aligned}
\dot{Z}&=H-H^* Z^2-\sigma^2 Z-
H^*\kappa\;,\\
\dot{\kappa}&=-4H^* Z\kappa-\sigma^2
(4\kappa+2Z^2)\;,\\
\dot{Y}&=F-F^* Y^2-\sigma^2 Y-
F^*\nu\;,\\
\dot{\nu}&=-4F^* Y\nu-\sigma^2
(4\nu+2Y^2)\;.
\end{aligned}
\label{eq:abr2}
\end{equation}
Solutions of this system are shown in Fig.~\ref{fig2}(b) with circles, for the same parameters as used in Fig.~\ref{fig2}(a), but with addition of a small noise $\sigma^2=10^{-4}$. This solution practically overlaps with the solution of the full equations \eqref{eq113} (solid blue line), where the infinite system \eqref{eq113} was truncated at a large number of modes $m=200$. This comparison confirms the quality of the cumulant approximation. Also we show in Fig.~\ref{fig2}(b) the OA solution for the noise-free case (the same dashed green line as in panel (a)). Importantly, the solution of system~\eqref{eq:abr2} is an attractor: we checked this by starting from different initial conditions in the (truncated as described above) full equations \eqref{eq113} (these initial conditions cannot be tested within system~\eqref{eq:abr2}, because it is valid only for small higher cumulants and not for generic initial states with potentially large higher cumulants). We illustrate convergence to the solution near the OA manifold described by system~\eqref{eq:abr2} in Fig.~\ref{fig2}(c).

With this example we see, that
noise acts in a stabilizing manner on the dynamics of the populations of identical oscillators. The probability density evolves toward a state close to the OA manifold. This state is well described, for small noise, by the first and the second circular cumulants. The distance to the OA manifold is visible even for small noise (cf.\ distance between the green curve and the circles in Fig.~\ref{fig2}(b)).

In conclusion, we have developed an analytic approach yielding closed equations for the collective modes for ensembles of noisy coupled phase oscillators. The equations generalize the Ott-Antonsen approach, valid in the noise-free situation, to the case of small noise. Our theory is based on the reformulation of the dynamics in terms of the circular cumulants. These new variables have a nice property: all high cumulants vanish on the OA manifold, thus providing a natural way to construct a perturbation procedure, using the noise intensity as a small parameter. Our equations account for the leading order in this parameter.

For the Abrams et al.\ chimera model, we demonstrated, that small noise makes a neighborhood of the OA manifold stable even for identical populations: a solution far from this manifold converges to a $\sigma^2$-vicinity of the OA manifold, where it can be described by the system derived in this letter. We expect this stabilizing effect of noise to be a rather generic property. However, a systematic analysis of different situations, especially of states far from the OA manifold, where a nontrivial interplay between noise and the deterministic dynamics may occur, is necessary to clarify the problem.

The method of circular cumulants can potentially be used to develop a perturbation approach for other situations, where the conditions of validity of the OA theory are slightly violated. These results will be reported elsewhere. At this point it is instructive to compare the cumulant approach of this paper with the perturbation theory developed in Ref.~\cite{Vlasov-Rosenblum-Pikovsky-16}. Theory~\cite{Vlasov-Rosenblum-Pikovsky-16} uses the Watanabe-Strogatz formalism and provides results in terms of corrections to the WS global variables. They are, however, different from the usual order parameters used in this paper, thus equations obtained here allow for a direct interpretation. Our approach is, however, restricted to the thermodynamic limit.

\acknowledgments
The authors thank V.\ Vlasov, M.\ Rosenblum, R.\ Mirollo, and J.\ Engelbrecht for fruitful discussions.
Work of A.P.\ on the Kuramoto model was supported by the Russian Science Foundation (grant No.\ 17-12-01534).
Work of L.S.K.\ and D.S.G.\ on general development of the cumulant approach was supported by the Russian Science Foundation (grant No.\ 14-12-00090).
The paper was finalized during the visit supported by G-RISC (grant No.\ M-2018a-7).


\begin{thebibliography}{36}
\bibitem[]{footnote1}%
  \BibitemOpen
  {The normalization
  of the circular cumulants $\varkappa_m$ differs from that of the conventional
  cumulants $K_m\!=\!(m\!-\!1)!\varkappa_m$}
  \BibitemShut{NoStop}%
\end{thebibliography}
%

\end{document}